\documentclass[12pt]{iopart}
\usepackage[dvips,final]{graphicx}
\usepackage{latexsym}

\usepackage{iopams}
\begin{document}

\newcommand{\bra}[1]{\ensuremath{\left\langle #1 \right\rvert}}
\newcommand{\ket}[1]{\ensuremath{\left\lvert #1 \right\rangle}}
\newcommand{\braket}[2]{\ensuremath{\left\langle #1 \vert #2 \right\rangle}}
\newcommand{\etion}[1]{\ensuremath{\left\langle #1 \right\rangle}}
\newcommand{\abs}[1]{\ensuremath{\left\lvert #1 \right\lvert}}
\newcommand{\norm}[1]{\ensuremath{\left\lVert #1 \right\rVert}}
\newcommand{\dbar}{{\mathchar'26\mkern-11mu\mathrm{d}}}
\newcommand{\hsbar}{{\mathchar'26\mkern-7mu\mathrm{h}}}
\renewcommand{\geq}{\geqslant}
\renewcommand{\leq}{\leqslant}
\newcommand{\bld}[1]{\boldsymbol{#1}}
\newcommand{\ub}{\boldsymbol{u}}
\newcommand{\xb}{\boldsymbol{x}}
\newcommand{\yb}{\boldsymbol{y}}
\newcommand{\kb}{\boldsymbol{k}}
\newcommand{\pb}{\boldsymbol{p}}
\newcommand{\rb}{\boldsymbol{r}}
\newcommand{\Rb}{\boldsymbol{R}}
\newcommand{\as}{\mathrm{a}}
\newcommand{\cs}{\mathrm{c}}
\newcommand{\ds}{\mathrm{d}}
\newcommand{\fs}{\mathrm{f}}
\newcommand{\gs}{\mathrm{g}}
\newcommand{\hs}{\mathrm{h}}
\newcommand{\is}{\mathrm{i}}
\newcommand{\ks}{\mathrm{k}}
\newcommand{\ps}{\mathrm{p}}
\newcommand{\rs}{\mathrm{r}}
\newcommand{\ssl}{\mathrm{s}}
\newcommand{\As}{\mathrm{A}}
\newcommand{\Ds}{\mathrm{D}}
\newcommand{\Fs}{\mathrm{F}}
\newcommand{\Is}{\mathrm{I}}
\newcommand{\Ns}{\mathrm{N}}
\newcommand{\Ss}{\mathrm{S}}
\newcommand{\Ts}{\mathrm{T}}
\newcommand{\Bcal}{\ensuremath{\mathcal{B}}}
\newcommand{\Dcal}{\ensuremath{\mathcal{D}}}
\newcommand{\Lcal}{\ensuremath{\mathcal{L}}}
\newcommand{\Mcal}{\ensuremath{\mathcal{M}}}
\newcommand{\Ncal}{\ensuremath{\mathcal{N}}}
\newcommand{\Ocal}{\ensuremath{\mathcal{O}}}
\newcommand{\Scal}{\ensuremath{\mathcal{S}}}
\newcommand{\Tcal}{\ensuremath{\mathcal{T}}}
\newcommand{\Wcal}{\ensuremath{\mathcal{W}}}
\newcommand{\ksl}{\slashed{k}}
\newcommand{\psl}{\slashed{p}}
\newcommand{\pog}[1]{\ensuremath{\mathsf{T}_{\gamma}\left( #1 \right)}}
\newcommand{\intf}[1]{\ensuremath{\int\sd^4 #1 \,}}
\newcommand{\intfb}[1]{\ensuremath{\int\dbar^4k \,}}
\newcommand{\inttb}[1]{\ensuremath{\int\dbar^3k \,}}
\newcommand{\tintfb}[2]{\ensuremath{\frac{\is}{\beta} \sum_{{#1}=-\infty}
    ^{\infty} \int\dbar^3{#2} \,}}
\newcommand{\dbd}[2]{\ensuremath{\frac{\ds #1}{\ds #2}}}
\newcommand{\pdbd}[2]{\ensuremath{\frac{\partial #1}{\partial #2}}}
\newcommand{\mbp}[1]{\ensuremath{\left(\frac{m^2}{4\uppi}\right)^{#1/2}}}
\newcommand{\bbol}[2]{\ensuremath{\frac{#1}{\e^{#2}-1}}}
\newcommand{\fbol}[2]{\ensuremath{\frac{#1}{\e^{#2}+1}}}
\newcommand{\omk}{\ensuremath{\omega_{\kb}}}
\newcommand{\head}[1]{{\textbf{\footnotesize #1}}}
\newcommand{\ph}[1]{\phantom{#1}}
\newcommand{\ix}[1]{\indices{#1}}

\newcommand{\be}{\begin{equation}}
\newcommand{\ee}{\end{equation}}
\newcommand{\bea}{\begin{eqnarray}}
\newcommand{\eea}{\end{eqnarray}}
\newcommand{\lag}{{\cal L}}
\newcommand{\vac}{{\rm vac}}
\newcommand{\leff}{\lambda_{\rm eff}}
\newcommand{\mpl}{M_{\rm P}}
\newcommand{\mn}{{\mu\nu}}
\newcommand{\nn}{\nonumber}
\newcommand{\gmn}{g_{\mu\nu}}
\newcommand{\Tmn}{T_{\mu\nu}}
\newcommand{\psibar}{\bar{\psi}}
\newcommand{\phibar}{\bar{\phi}}
\newcommand{\rhobar}{\bar{\rho}}
\newcommand{\Ubar}{\bar{U}}
\newcommand{\gt}{\tilde{g}}
\newcommand{\sg}{\sqrt{-g}\,}
\newcommand{\sgt}{\sqrt{-\tilde{g}}\,}
\newcommand{\m}{{\rm (m)}}
\newcommand{\barphi}{\frac{\phi}{\mpl}}
\newcommand{\ut}{{\widetilde U}}
\newcommand{\vt}{{\widetilde V}}
\newcommand{\lz}{{\widetilde \mu}}
\def\note#1{{\bf *** #1 ***}}

\renewcommand{\Im}{\mathrm{Im}}     
\renewcommand{\Re}{\mathrm{Re}}

\unitlength=1mm

\title{Modified-Source Gravity and Cosmological Structure Formation}

\author{Sean M. Carroll$^{1}$, Ignacy Sawicki$^{1}$,
Alessandra Silvestri$^{2}$ and Mark Trodden$^{2}$}

\address{$^{1}$ Enrico Fermi Institute, Department of Physics,
and Kavli Institute for Cosmological Physics, University of Chicago,
5640 S. Ellis Avenue, Chicago, IL 60637. \\
{\tt carroll@theory.uchicago.edu, sawickii@theory.uchicago.edu}}
\address{$^{2}$ Department of Physics, Syracuse University,
Syracuse, NY 13244-1130, USA.\\
{\tt asilvest@physics.syr.edu, trodden@physics.syr.edu}}

\begin{abstract}
One way to account for the acceleration of the universe is to modify
general relativity, rather than introducing dark energy.  Typically, 
such modifications introduce new degrees of
freedom.  It is interesting to consider models with no new degrees
of freedom, but with a modified dependence on the conventional
energy-momentum tensor; the Palatini formulation of $f(R)$ theories
is one example.  Such theories offer an interesting testing ground
for investigations of cosmological modified gravity.  In this paper
we study the evolution of structure in these
``modified-source gravity'' theories.
In the linear regime, density perturbations exhibit scale
dependent runaway growth at late times and, in particular, a mode of
a given wavenumber goes nonlinear at a higher redshift than in the
standard $\Lambda$CDM model. We discuss the implications of this
behavior and why there are reasons to expect that the growth will be
cut off in the nonlinear regime. Assuming that this holds in a full
nonlinear analysis, we briefly describe how upcoming measurements
may probe the differences between the modified theory and the
standard $\Lambda$CDM model.
\end{abstract}
\maketitle


\section{Introduction}
\label{intro}
The concordance cosmological model describes a universe
consisting of approximately 5\% ordinary matter, 25\% dark matter, and
70\% dark energy.  While it fits a wide variety of data, this model
relies heavily on the existence of a ``dark sector'' comprising 95\%
of the universe.  Given the mysterious nature of dark matter and dark
energy, and the fact that their existence is inferred exclusively
through their gravitational effects, it is natural to wonder whether
the apparent need for these components could be a sign that gravity
is deviating from conventional general relativity (GR) on large scales.

Dynamical measurements that are taken to imply the existence of dark
matter generally refer to length scales of kiloparsecs or greater,
while evidence for dark energy comes from the acceleration of the
universe, a phenomenon characteristic of the present Hubble radius of
order one gigaparsec.  The most precise experimental tests of GR, meanwhile,
probe much smaller length scales \cite{Will:2001mx};
Solar System measurements cover less than a milliparsec, while the
binary pulsar PSR~1913+16 has an orbital semi-major axis less than a
microparsec.  We can therefore imagine that there is a large dynamical
range over which gravity obeys Einstein's equation to high precision,
while behaving differently at sufficiently large scales.

Within the modified-gravity category there have been several
different proposals: induced gravity in a class of extra dimensional
models (DGP braneworlds) \cite{Dvali:2000hr,Deffayet:2000uy,Deffayet:2001pu};
phenomenological modifications of the Friedmann
equation of cosmology \cite{Freese:2002sq,Dvali:2003rk};
and direct, manifestly covariant
modifications of the four-dimensional Einstein-Hilbert
action~\cite{Carroll:2003wy,Capozziello:2003tk,Vollick:2003aw,
Flanagan:2003rb,Flanagan:2003iw,Vollick:2003ic,Soussa:2003re,
Nojiri:2003ni,Carroll:2004de}.
(For some other ideas see
\cite{Arkani-Hamed:2003uy,Gabadadze:2003ck,Moffat:2004nw,Clifton:2004st}.)
In the last of these approaches, the simplest models involve an action
obtained by integrating a function of the curvature scalar $R$,
\be
  S = \int d^4x \sg f(R)\,,
  \label{fofr}
\ee
and varying with respect to the metric $g_\mn$. These models were
soon discovered to be dual to scalar-tensor theories, which lead to
unacceptable deviations from GR in the solar system
\cite{Chiba:2003ir}. However, certain models based on inverse powers
of more general curvature invariants \cite{Carroll:2004de}
(particularly those involving the square of the Riemann tensor) have
recently been shown to agree with solar system
tests~\cite{Navarro:2005gh} and to provide a good fit to the
supernovae data~\cite{Mena:2005ta}, although tight constraints arise
from the requirement that they be
ghost-free~\cite{Navarro:2005da,DeFelice:2006pg,Calcagni:2006ye}.

On theoretical grounds, however, there are obstacles to constructing
infrared modifications of GR that will escape detection on smaller scales.
In a weak-field expansion around flat spacetime,
GR describes the propagation of massless spin-2 gravitons coupled to the
energy-momentum tensor $T_{\mu\nu}$, including that of the gravitons
themselves.  But such a theory is essentially unique; it has long been
known that we can start with a field theory describing a
transverse-traceless symmetric two-index tensor propagating in Minkowski
space, couple it to the energy-momentum tensor of itself and other
fields, and demonstrate iteratively that the background metric
disappears, leaving instead a curved metric obeying Einstein's equation.
It is therefore generally believed that infrared modifications
of GR will necessarily involve the introduction of new degrees of
freedom.

For the purpose of explaining the acceleration of the universe, however,
there is a loophole in this argument.
The properties of gravitons define both the propagation of linearized
gravitational waves and the Coulomb form of the Newtonian gravitational
potential in the weak-field limit.
But there is more to gravity than gravitons, even in the infrared.
In particular, the evolution of a Robertson-Walker cosmology
is described by the Friedmann equation,
\be
  H^2 = \frac{8\pi G}{3}\rho -\frac{\kappa}{a^2}\,,
  \label{friedmann}
\ee
where $\rho$ is the energy density, $\kappa$ is the spatial curvature,
and $H$ is the Hubble parameter,
related to the cosmological scale factor $a(t)$ by $H=\dot{a}/a$.
The Friedmann equation has nothing to do with gravitons; it is a
constraint, relating the instantaneous expansion rate to the energy
density and curvature.  In fact, the Friedmann equation is a particular
example of the Hamiltonian constraint of general relativity, relating
the embedding of a partial Cauchy surface in spacetime to its
intrinsic geometry and the energy-momentum tensor.

Because the Friedmann equation arises as a consequence of the constraint
structure of GR rather than from the behavior of gravitons,
we are free to imagine modifying it without introducing new degrees
of freedom.  In particular, Einstein's equation relates the Einstein
tensor to the energy-momentum tensor for matter,
\be
  G_\mn = 8\pi G T_\mn\,.
  \label{einstein}
\ee
In a theory describing a set of matter fields $\chi_i$ with an
action $S_\m[g_\mn, \chi_i]$, the energy-momentum tensor is
defined by
\be
  T_\mn \equiv -\frac{2}{\sg}\frac{\delta S_\m}{\delta g^\mn}\,.
  \label{tmunu}
\ee
We may ask, then, to what other sources could gravitons couple?
In other words, we might define an alternative theory obeying
\be
  G_\mn = 8\pi G \tau_\mn(\chi_i)\,,
  \label{fmunu}
\ee
where $\tau_\mn$ is some symmetric $(0,2)$ tensor that is conserved
by virtue of the matter equations of motion, but differs from
the conventional form (\ref{tmunu}).  Since we have altered the
right-hand side of Einstein's equation without introducing any
new degrees of freedom, it may be possible to account for the
acceleration of the universe without dark energy while remaining
consistent with conventional tests of GR.

In fact this idea has been realized in the form of the Palatini
formulation of $f(R)$ gravity.  As Flanagan has shown
\cite{Flanagan:2003rb}, the equations of motion obtained by
separately varying the metric and connection in an action of the
form (\ref{fofr}) actually leads to a theory with \emph{fewer}
degrees of freedom; there is no propagating scalar, only the
massless spin-two graviton of ordinary GR. In this paper we further
consider models of this type, leaving behind the inspiration of the
Palatini formulation of $f(R)$ models and working directly with
theories that have no propagating scalar, which we dub
``modified-source gravity'' (MSG).  The theories we consider are not
precisely identical to the Palatini models, since we do not
separately vary the metric and connection, so that the equations for
the matter sector will be slightly different; nevertheless, there is
an essential similarity between the two ideas.

Flanagan has argued that models
of this type are experimentally excluded, as the effective matter
action includes higher-order couplings not seen in particle-physics
experiments.  However, as we argue below, there are a number of
reasons to believe that these constraints can be avoided; furthermore,
the MSG field equations are an interesting toy model of cosmological
modified gravity against which observations may be compared.

The possibility of distinguishing between modified gravity and dark
energy by comparing the expansion history of the universe to the
growth rate of cosmological perturbations has recently been
emphasized from a variety of approaches
\cite{Lue:2004rj,Knox:2005rg,Ishak:2005zs,Linder:2005in,Lue:2005ya,
Sawicki:2005cc,Koyama:2005kd,Koyama:2006ef,Song:2006sa,Bertschinger:2006aw,
Yamamoto:2006yv,Uzan:2006mf,Sawicki:2006jj,Song:2006jk}. Given the
paucity of viable alternative theories of cosmological gravity, it
is important to develop a model-independent perspective on the ways
in which this distinction can manifest itself; we know of no better
way to develop such intuition than to examine as many models as
possible.

In this paper we therefore explore the cosmology of modified-source
gravity by studying cosmological perturbation theory and extracting
predictions for the growth of large-scale structure.  We find that
there is a substantial boost in the growth of the gravitational
potential in comparison with ordinary $\Lambda$CDM. We discuss the
implications of this behavior and why there are reasons to expect
that the growth will cease in the nonlinear regime. Assuming that
this holds in a full nonlinear analysis, we briefly describe how
upcoming measurements may probe the differences between the modified
theory and the standard $\Lambda$CDM model.

Modified-source gravity is examined very much in the spirit of a toy
model, as a laboratory to help understand the possible differences
between theories of dark energy and theories that alter general
relativity.  It does not attempt
to solve the cosmological constant problem (we set the vacuum
energy to zero by hand), nor does it require less fine-tuning
than any other typical model of dark energy (we choose the
potential delicately so that acceleration happens at very late
times), or is there any obvious way in which it would naturally appear
as the low-energy limit of a more complete theory in the ultraviolet.
Nevertheless, it is sufficiently difficult to find
cosmological alternatives to general relativity that a simple explicit
model can be useful in helping to focus efforts to distinguish
between modified gravity and sources of dynamical dark energy.


\section{Modified-Source Gravity}
\label{msg}

\subsection{Scalar-tensor and $f(R)$ theories}
\label{stfr}

The Einstein-Hilbert action for general relativity
is
\be
  S_{\rm EH} = \int \ds^4x \sg \left(\frac{1}{2}\mpl^2R\right)  \,,
  \label{einsteinhilbert}
\ee
where $R$ is the curvature scalar and $\mpl = 1/\sqrt{8\pi G}$
is the reduced Planck mass.  Matter fields couple minimally to the
metric $g_\mn$.  The propagating degrees of
freedom of this theory are the two polarization
states of a massless spin-two graviton.

One of the simplest ways to modify GR without introducing
new fields is to consider actions that depend non-linearly on
$R$,
\be
  S_f = \int \ds^4x \sg f(R)  \,.
  \label{fofr2}
\ee
A particular example in which deviations from GR become important
at small curvatures is $f(R) = (\mpl^2/2)(R-\mu^4/R)$
\cite{Carroll:2003wy,Capozziello:2003tk}.
Interestingly, however, the linear Einstein-Hilbert term
(\ref{einsteinhilbert}) is the {\it only} function of $R$
that propagates a spin-2 graviton by itself; any other function
$f(R)$ gives rise to a scalar-tensor theory, with both a
spin-2 graviton and a spin-0 scalar \cite{tess,Magnano:1993bd,wands}.
This can be seen explicitly by considering an action with
gravity coupled to a scalar $\lambda$,
\be
  S = \int \ds^4x \sg \left[(R-\lambda)f'(\lambda)
  + f(\lambda)\right]\,,
  \label{lambdaaction}
\ee
where $f'(\lambda)=df/d\lambda$.  The field $\lambda$ functions as a
Lagrange multiplier, whose equation of motion sets $\lambda = R$.
Plugging this back into (\ref{lambdaaction}) yields (\ref{fofr2}),
so long as $f'' \neq 0$ (that is, if $f$ is anything other than linear).
Defining a new scalar field $\psi$ via
\be
  \psi = \frac{1}{2} \ln[f'(\lambda)]\,,
  \label{psilambda}
\ee
we can perform a conformal transformation of the form
\be
  \gt_\mn = \e^{2\psi}g_\mn\,.
\ee
The conformally-transformed action becomes that of ordinary general
relativity coupled to a propagating scalar field,
\be
  \widetilde{S} =  \int \ds^4x \sg \left[\frac{\mpl^2}{2} \widetilde{R}
  - 3\gt^\mn (\nabla\psi)^2 - \ut(\psi)\right]\,,
  \label{scalartensoraction}
\ee
where $(\nabla\psi)^2 \equiv g^\mn \nabla_\mu\psi \nabla_\nu\psi$.
This is the Einstein-frame version of the model, in which the Ricci
curvature appears by itself, but matter fields couple to the scalar
$\psi$ through the combination $g_\mn=\e^{-2\psi}\gt_\mn$; in the
matter frame given by (\ref{fofr2}), there is no direct coupling
between $\psi$ and the matter fields, and free particles move along
geodesics of $g_\mn$.  The potential is given by
\be
  \ut(\psi) = \frac{\lambda f'(\lambda)
  - f(\lambda)}{2f'(\lambda)}\mpl^2\,.
\ee
In this expression, $\lambda$ is taken to be a function of $\psi$ via
inverting (\ref{psilambda}).
(Note that our notation differs from that in \cite{Carroll:2003wy}.)

The problem with such a model is that, if we choose the function
$f(R)$ so that gravity is modified at small curvatures and the
universe accelerates at late times, the scalar field will be very
light, and the theory comes into conflict with Solar-System tests of
gravity.  In terms of the Brans-Dicke parameter $\omega$, the model
of \cite{Carroll:2003wy} is nearly equivalent (apart from the small
and presumably negligible potential) to a theory with $\omega=0$;
meanwhile, the best current limits come from measurements of the
Shapiro time delay from the Cassini mission, and give $\omega >
40,000$ \cite{Bertotti:2003rm}. However, recently some authors have
claimed that the GR limit of the dynamical equivalence between $f(R)$ and
scalar-tensor theories is not well behaved, since $f''(R)
\rightarrow 0$, and some $f(R)$ theories behave in the Solar System
in a manner perfectly consistent with current experimental limits
\cite{Faraoni:2006hx,Capozziello:2005bu,Capozziello:2006jj}.

Different strategies for avoiding this constraint have been
explored. One approach is to consider models with inverse powers of
other curvature invariants such as $R_\mn R^\mn$ and
$R_{\mn\rho\sigma}R^{\mn\rho\sigma}$
\cite{Carroll:2004de,Navarro:2005gh,
Mena:2005ta,Navarro:2005da,DeFelice:2006pg,Calcagni:2006ye}. As
mentioned in the introduction, the degrees of freedom of such models
differ from those of simple $f(R)$ theories, and may be consistent
with solar system tests of gravity. Alternatively, one may consider
the Palatini versions of $f(R)$ theories, in which the metric and
connection are treated as independent variables
\cite{Vollick:2003aw,Flanagan:2003rb,
Flanagan:2003iw,Vollick:2003ic,Sotiriou:2006qn}.  Unlike in the case
of the Einstein-Hilbert action, $f(R)$ theories do not have
identical equations of motion in the Palatini formulation as in the
conventional scenario based on the metric alone.  Indeed, Flanagan
has shown that the scalar degree of freedom disappears entirely
\cite{Flanagan:2003rb}. In the next
section we explore a possibility of this form, not through the
Palatini variation of an $f(R)$ action, but simply by eliminating
the scalar kinetic term by hand (although of course the formulations
are related).

\subsection{Eliminating the scalar}
\label{erasing}

Our goal is to take the scalar-tensor action (\ref{scalartensoraction}),
equivalent to (\ref{fofr2}), and eliminate the propagating scalar
degree of freedom.  We choose the most brutally direct approach:
simply erasing the kinetic term $-3\gt^\mn \nabla_\mu\psi
\nabla_\nu\psi$ from (\ref{scalartensoraction}).  We are left with a
new theory in which $\psi$ is a Lagrange multiplier, without any
dynamics of its own. Alternatively,
we could imagine multiplying the kinetic term by a constant, and
then taking the limit as the constant went to zero;  in that limit,
the scalar is still propagating, but decouples from any other fields.
Generally, radiative corrections would tend to drive this constant
to order unity.  We will nevertheless set it to zero for purposes of
this paper, accepting this as one of the fine-tunings that inevitably
accompanies models of dynamical dark energy.\footnote{Of course,
the kinetic term may always be
canonically normalized by a field redefinition.  Setting the coefficient
of the kinetic term to zero is therefore equivalent to taking the potential
and the couplings all simultaneously to infinity.}

The full Einstein-frame action for the theory, including additional
matter fields $\chi_i$, is
\be
  \widetilde{S} = \int \ds^4x
  \sgt\left[{\mpl^2\over 2}\widetilde{R}
  - \ut(\psi)\right]
  + \widetilde{S}_\m[\e^{-2\psi}\gt_\mn, \chi_i]\ .
  \label{lag2e}
\ee 
In the absence of the matter action, this model would be
completely trivial. The scalar is non-propagating, and its equation
of motion would fix $\psi$ at any allowed value $\psi_0$ at which
$d\ut/d\psi=0$. The model is then simply general relativity with a
vacuum energy given by $\ut(\psi_0)$.  The coupling to matter,
however, leads to important consequences.  We can undo the conformal
transformation to return to the matter-frame metric $g_\mn =
\e^{-2\psi}\gt_\mn$, yielding \be
  S = \int \ds^4x \sg\left[{\mpl^2\over 2}\e^{2\psi} R + 3
  \e^{2\psi}(\nabla\psi)^2
  - U(\psi)\right]+ S_\m[g_\mn, \chi_i]\ ,
  \label{lag2}
\ee
where the new potential is
\be
  U(\psi)= \e^{4\psi}\ut(\psi)\ .
  \label{newpotential}
\ee

The actions (\ref{lag2}) or (\ref{lag2e}) define modified-source
gravity. We have left behind our original inspiration from $f(R)$
gravity, so the potential $U(\psi)$ is now simply a free function
that defines the theory. In the form (\ref{lag2}), the model
resembles a conventional scalar-tensor theory, with a kinetic term
for $\psi$ and a direct coupling to the curvature scalar.  However,
there are implicitly derivatives of $\psi$ in the ${1\over 2}
\e^{2\psi}R$ term, which would be revealed after integration by
parts, and would cancel the explicit kinetic term.  The scalar field
$\psi$ is actually completely non-dynamical, as is evident from the
Einstein-frame expression (\ref{lag2e}), in which no derivatives of
$\psi$ appear.  In principle, this field could be integrated out
exactly, and we will examine this approach in the next section; for
the moment, it is more convenient to leave $\psi$ explicitly in the
action and field equations.

Let's examine the theory in the matter frame (\ref{lag2}).
The gravity equation of motion, obtained by varying with respect to
$g_\mn$, can be written
\be
  \e^{2\psi}\mpl^2 G_\mn = T_\mn^{({\rm m})} + T_\mn^{(\psi)}\,.
  \label{einsteineq0}
\ee
Here, $T_\mn^{(\psi)}$ is the effective energy-momentum tensor for
$\psi$,
\be
  T_\mn^{(\psi)} =
  - \left[U(\psi) + (\nabla\psi)^2 + 2 \Box\psi \right]g_\mn
  - 2\nabla_\mu\psi\nabla_\nu\psi + 2\nabla_\mu\nabla_\nu\psi\,,
  \label{tmunupsi}
\ee
where $\Box \equiv g^\mn\nabla_\mu\nabla_\nu$.
The equation of motion for $\psi$ is
\be
  \Box\psi + (\nabla\psi)^2 + {1\over 6\mpl^2}e^{-2\psi}\dbd{U}{\psi}
  - {1\over 6}R = 0\,.
  \label{msgscalar}
\ee
Again, these equations bear a close resemblance to those of ordinary
scalar-tensor gravity.  However, we can take the trace of
(\ref{einsteineq0}) to obtain
\be
  R = {\e^{-2\psi}\over \mpl^2}\left[-T + 4U(\psi)\right] + 6(\nabla\psi)^2
  + 6\Box\psi\,,
\ee where $T\equiv g^\mn T_\mn^{({\rm m})}$ is the trace of the
matter energy-momentum tensor alone. Subtracting this from the
scalar equation (\ref{msgscalar}) leaves \be
  \dbd{U}{\psi} - 4U(\psi) = -T\,.
  \label{constraint0}
\ee
Thus, the trace of the gravity equation can be used to eliminate all
spacetime derivatives from the scalar equation, leaving us with a
{\it purely algebraic equation for $\psi$ in terms of the matter
fields}.

If we start with $U(\psi)$, we could use this equation to get
$T(\psi)$, and invert that to get $\psi(T)$.
If we start with $\psi(T)$ [or equivalently $T(\psi)$], we can
then express the potential in terms of $\psi$ as
\be
  U(\psi) = -\e^{4\psi} \int_{\psi_0}^\psi
  \e^{-4\psi'}T(\psi')\, \ds\psi'\,.
  \label{Upsi}
\ee
Lastly, if we know $U(T)$ instead of $\psi(T)$, we could get $\psi$ via
\be
  \psi(T) = -\int \frac{1}{T}\left(\dbd{U}{T}
  + 4U\right)\ds T\,.
  \label{psiT}
\ee


\section{Comparison with Experiment}
\label{experiment}

\subsection{Solar-System tests}
\label{solarsystem}

Consider the vacuum equations, where $T_\mn = 0$.  Then from
(\ref{constraint0}) the scalar is pinned at some value $\psi_0$, with
\be
  4U(\psi_0) - U'(\psi_0) = 0\ .
\ee
Looking back at (\ref{einsteineq0}), the vacuum gravitational field equation
is
\be
  \mpl^{2}G_\mn = - \e^{-2\psi} U(\psi_0)g_\mn\ .
  \label{vacuum}
\ee
Thus, $\e^{-2\psi_0} U(\psi_0)$ plays the role of an ordinary
cosmological constant.  Otherwise, the vacuum equation is {\em
precisely} the vacuum Einstein equation.  The Schwarzschild metric
is therefore an exact solution of this theory (if the
vacuum energy vanishes; otherwise it would be Schwarzschild-de~Sitter).
Gravitational waves in free space are completely identical to those
in general relativity; the propagating degrees of freedom are simply
those of a massless spin-2 graviton.

In $f(R)$ theories, Schwarzschild (for which $R_\mn=0$)
is generally also an exact solution
to the vacuum field equations, even though those equations are not
identical to Einstein's.  However, it is not a \emph{unique} solution,
even with spherical symmetry; Birkhoff's theorem, which relies on
Einstein's equation, fails to apply.  This failure
can ultimately be traced to the existence of new degrees of freedom;
gravitating bodies are not only characterized by their mass, but also by a
charge that couples to a new scalar field. Modified-source gravity, in
contrast, has no new degrees of freedom, and Einstein's equation holds
in vacuum, so Birkhoff's theorem applies.  The Solar-System tests of
gravity that rule out simple $f(R)$ theories are completely compatible
with MSG.

We are free to choose the ``potential'' $U(\psi)$ to satisfy phenomenological
requirements.  If our interest is in causing the universe to accelerate
at late times without affecting local tests of gravity, these
include the following:
\begin{itemize}
\item For energy densities $\rho$ larger than some critical value
$\rho_*$ of order the present mass density of the universe, we want
to recover conventional cosmological expansion ($a\propto t^{2/n}$ when
the total energy density is evolving as $\rho \propto a^{-n}$).  This
is achieved if $\psi\rightarrow \psi_* = {\rm constant}$
at $\rho \geq \rho_*$.
For constant $\psi$, the additional energy-momentum contributions
from (\ref{tmunupsi}) are simply those of a constant effective vacuum energy
$U(\psi) < \rho_*$.  In particular, the effective gravitational constant
will be constant, and we recover conventional
Friedmann-Robertson-Walker cosmology.
\item Since $\rho_*$ is smaller than the density of any compact
astrophysical object, the previous requirement guarantees that
we will automatically get
$\psi = \psi_*$ inside planets and stars.  The strength of gravity will
not be significantly density-dependent, so the predictions of Solar-System
tests or the binary pulsar are unaffected without further restrictions.
\item For $\rho \leq \rho_*$ the universe should accelerate.
In section~\ref{amodel} we consider one concrete choice of $U(\psi)$
that accomplishes this goal.  We might also ask that the value of $U$
vanish when $\rho=0$, so that Minkowski space is a solution to the theory;
otherwise, late-time acceleration could simply be blamed on the new
potential energy, which wouldn't really be different from ordinary dark
energy.
\end{itemize}

\subsection{Matter interactions}
\label{matterinteractions}

Despite the characterization of MSG as a theory of ``gravity,'' we
can certainly imagine integrating out the scalar field in the
Einstein-frame action (\ref{lag2e}) to obtain a
modified matter Lagrangian coupled to general relativity.  
The result will be an action written purely in
terms of the ordinary matter fields and gravity; if the matter
action looks conventional in the matter frame, it will generally be
an ungainly mess in the Einstein frame with $\psi$ integrated out.
As noted by Flanagan \cite{Flanagan:2003rb}, this implies the
existence of higher-order (nonrenormalizable) terms in the matter
action, which will be proportional to inverse powers of a very small
mass scale $\mu$ characterizing the regime in which gravity is to
be modified.  Since $\mu$ should be very low in models 
relevant to late-time acceleration, such new interactions would
presumably be easily noticeable in experiments (and, needless to
say, have not been seen).

In addition to the introduction of new nonrenormalizable interactions,
another phenomenological problem identified by Flanagan
\cite{Flanagan:2003rb} relates to the nonlinearity of the
source for gravity in MSG.  The effective energy
density is a nonlinear function of the conventional matter-frame
energy-momentum tensor.  Thus, it is illegitimate to average
the energy density of a particulate medium over some coarse-graining
to obtain a fluid description.  In the limit where a body is made
of pointlike particles, the density is infinite at the locations of
the particles and zero in between, and treating it as a smooth
density profile would be a mistake.  

Despite these reasonable concerns, for the rest of this paper we will
imagine that MSG is experimentally viable and an averaged fluid
description for the matter fields is sufficient.  Our primary 
justification for sidestepping this important issue is the dramatic
separation of scales between the regimes we are considering (evolution
of cosmological perturbations) and those in which the above considerations
become relevant (atomic and particle-physics scales).  We therefore
find it interesting to consider the MSG equations as a phenomenological 
description of gravity in the infrared, even if we do not have
a viable model on small scales.  In order to best understand how we
can observationally distinguish dark energy from modified
gravity, it is important to characterize different ways in which 
modified gravity could manifest itself cosmologically.  Since the MSG
equations provide a consistent dynamical description of gravity on
very large scales, studying their observational consequences is
a useful practical exercise.

Furthermore, it is certainly possible to imagine ways in which the problems of
higher-order interactions and the non-linear energy-momentum tensor
could be overcome.  Perhaps the most straightforward would be to 
imagine that the kinetic term for our auxiliary scalar $\psi$ 
were not exactly zero, but
rather some very small number; the fluctuations of this field could then
serve to average over the density of particles on sufficiently
small scales.  Alternatively, if the dark matter is some 
smoothly-distributed bosonic condensate (such as axions), the fluid
description used in this paper could be completely accurate.  In
fact, it is important to note that experimental results relevant to
the question of higher-order interactions do not actually take place
in a true vacuum, but rather in the presence of whatever dark matter
background exists in the Solar System.  Since the local density of dark
matter is likely to be substantially
greater than that of the average density of the universe, terrestrial
experiments take place in a regime where gravity is accurately described
by GR and novel MSG effects are suppressed.  Finally, 
Flanagan also notes \cite{Flanagan:2003iw} that the new
interactions are strongly-coupled in the low-energy regime in which
they are purportedly observable.  All of these possibilities are 
interesting and subtle, and deserve greater attention than we will 
give them in this paper, where our interest is exclusively cosmological;
further development of the theory to put it on more secure
microphysical foundations would be very worthwhile.


\section{Robertson-Walker Cosmology}
\label{cosmology}

\subsection{The Modified Friedmann equation}
\label{friedmannequation}

In this subsection we examine the evolution of
a homogeneous and isotropic universe, deriving the modified Friedmann
equation for MSG.  In the next subsection
we focus on an explicit choice for the potential $U(\psi)$
that leads to late-time acceleration.

Consider a flat Robertson-Walker metric,
\be
  \ds s^2 = -\ds t^2 + a^2(t) \left[\frac{\ds r^2}{1-\kappa r^2} + r^2(\ds\theta^2
  + \sin^2\theta \ds\phi^2)\right]\,,
\ee
where $\kappa$ is zero, positive or negative depending on the
curvature of the spatial hypersurface.  It is not normalized, as we have
chosen instead
to set $a=1$ at the present time.  The matter fields are taken to be a
perfect fluid, with energy-momentum tensor
$T^\mu{}_\nu = {\rm diag}(-\rho, p,p,p)$.
The 00 component of the gravitational field equation (\ref{einsteineq0})
becomes
\be
  3H^2 + 3\frac{\kappa}{a^2} = {\e^{-2\psi}\over \mpl^2}
  \left[\rho + U(\psi)\right]
  - 3{\dot\psi}^2 - 6{{\dot a} \over a}\dot\psi\,,
\ee
where $H={\dot a}/a$.
We can convert the time derivatives of $\psi$ into derivatives with
respect to $a$ by using
\be
  \dot\psi = {\dot a} \dbd{\psi}{a}
  = H\left(\dbd{\psi}{\ln{a}}\right)\,.
\ee
Collecting terms proportional to $H^2$ and dividing, we obtain
\begin{equation}
  H^2 = \left(1+\dbd{\psi}{\ln a}\right)^{-2}\left[
  \frac{\e^{-2\psi}}{3M_{\rm P}^2}[\rho+U(\psi)]-\frac{\kappa}{a^2}\right]
  \,, \label{e:fried}
\end{equation}
which serves as the cosmological evolution equation for MSG.

There are two obvious modifications from the conventional Friedmann
equation: a potential-energy
contribution $U(\psi)$, and a variable-strength effective Newton's constant,
\be
  8\pi G_{\rm eff} = \frac{\e^{-2\psi}}{\mpl^2}
  \left(1+\dbd{\psi}{\ln a}\right)^{-2}\,.
\ee
When $\kappa =0$ and $\rho \gg U$, this is the quantity that
relates the energy density to the expansion rate.
In fact, both the potential and the effective Newton's
constant are simply functions of the density $\rho$
and the pressure $p$ through the $\psi$ equation (\ref{constraint0}).
The new equation is therefore reminiscent of the Cardassian model
\cite{Freese:2002sq}, in which the right-hand side of the Friedmann
equation is a non-linear function of $\rho$.

\subsection{A model}
\label{amodel}

There is a great amount of freedom in the choice of MSG dynamics,
represented by the arbitrary function $U(\psi)$. In this subsection
we discuss one simple example that satisfies the criteria laid out
in section~\ref{solarsystem}, including late-time acceleration.

For simplicity we imagine that the matter is represented
by a pressureless fluid, so that $\psi$ can be thought of as a
function of the matter energy density
\be
\rho = -T = \rho_0 a^{-3}\, ,
\ee
where $\rho_0$ is the average cosmological energy density today.
Given the relations (\ref{constraint0}), (\ref{Upsi}), and (\ref{psiT}),
we can choose to specify any of the three quantities $\psi$,
$U$, and $\rho$ as a function of any one of the others, and
their relationship to the third will be automatically determined.

We would like to choose behavior for which $\psi$ is approximately
constant when the energy density is substantially higher than the
present average cosmological density $\rho_0$, and \emph{decreases}
for $\rho \ll \rho_0$.  A convenient form to choose is
\be
    \e^{-4\psi}=\alpha\left(\frac{\rho_{0}}{\rho}\right)+1 \, ,
    \label{model2}
\ee
with $\alpha$ a dimensionless parameter defining the model.  The
limiting behavior is given by
\be
  \psi(\rho\rightarrow\infty) = 0\ , \qquad
  \psi(\rho \rightarrow 0) = \frac{1}{4}\ln{(\rho)} \rightarrow -\infty,
\ee
satisfying the aforementioned criteria.

\begin{figure}[ht]
\begin{centering}
\includegraphics[width=\columnwidth]{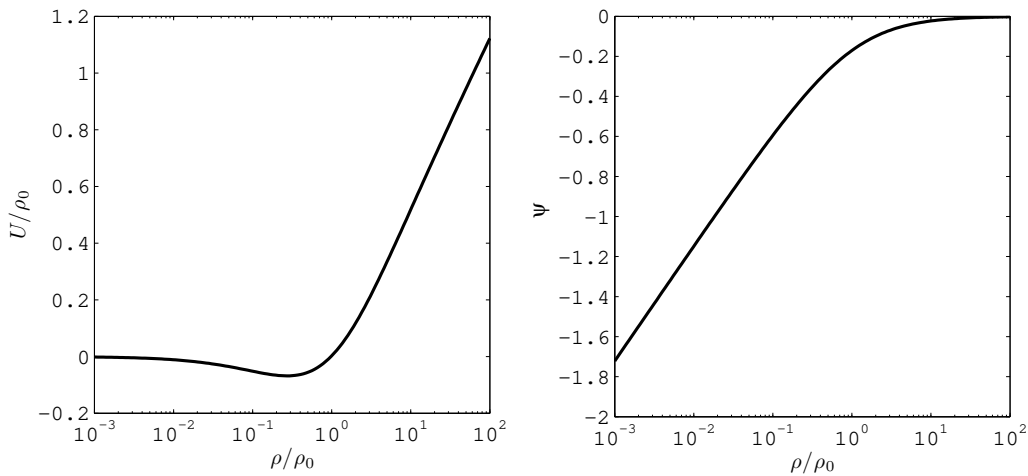}
\caption{The potential $U$ (in units of $\rho_0$) and scalar field $\psi$
(in units of $\mpl$) as
functions of the cosmological energy density $\rho$.}
\label{rhoplots}
\end{centering}
\end{figure}

The convenience of this model arises from our ability to
analytically determine the potential as a function of the density,
\be
  U(\rho) = -\frac{\alpha \rho_0 \rho}{4(\alpha\rho_0
  + \rho)}\ln\left(\alpha \frac{\rho_0}{\rho}\right)\,,
\ee
or equivalently as a function of $\psi$,
\be
  U(\psi) = \alpha\rho_0 \e^{4\psi}\left[\psi - \frac{1}{4}
  \ln\left(1-\e^{4\psi}\right)\right]\,.
\ee
As $\rho\rightarrow 0$ we get
\be
  U(\rho \rightarrow 0) = \frac{\rho}{4} \ln\left(\frac{\rho}{\rho_0}\right)
   \rightarrow 0\,.
\ee
The fact that the potential vanishes at zero density guarantees
that Minkowski space is a solution to the model.
At large density we have
\be
  U(\rho \rightarrow \infty) = \frac{\alpha\rho_0}{4}
  \ln\left(\frac{\rho}{\rho_0}\right) \,,
\ee
which rises more slowly than $\rho$, so that the potential does not
dominate at early times.  The behavior of $U$ and $\psi$ as functions
of $\rho$ is shown in Figure~\ref{rhoplots}.

The other factor appearing in the Friedmann equation (\ref{e:fried})
involves $d\psi/d\ln{a}$, which can be written in this model as
\be
  \dbd{\psi}{\ln{a}} = -3\rho \dbd{\psi}{\rho} =
  -\frac{3\alpha\rho_0}{4(\alpha\rho_0 + \rho)}\,.
\ee
Putting everything together, the full Friedmann equation (\ref{e:fried})
for the specific model defined by (\ref{model2}) becomes
\be
  H^2 = \left(\frac{4\alpha\rho_0 + 4\rho}
  {\alpha\rho_0 + 4\rho}\right)^2 \left[\frac{1}{3\mpl^2}
  \left(\frac{\alpha\rho_0 + \rho - (\alpha\rho_0/4)\ln(\alpha\rho_0/\rho)}
  {\sqrt{\alpha\rho_0 + \rho}}\right)\sqrt{\rho} - \frac{\kappa}{a^2}
  \right].
  \label{e:fried2}
\ee
It seems unlikely that such an expression would have been guessed at
as a phenomenological starting point for exploring cosmological dynamics
in theories of modified gravity.

\begin{figure}[ht]
\begin{centering}
\includegraphics[width=\columnwidth]{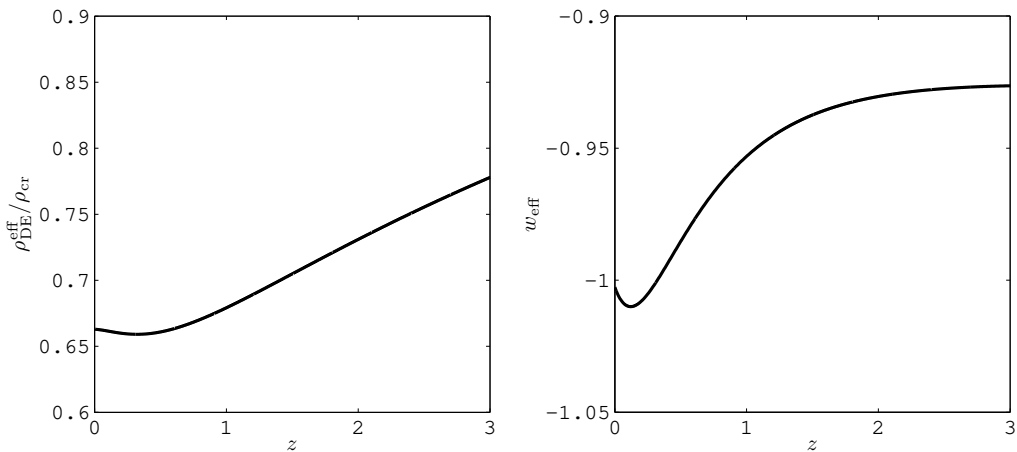}
\caption{The plot on the left shows the effective dark-energy
density, in units of the matter density today, that observers would
reconstruct.  (That is, the dark-energy density that would lead to
equivalent behavior for the scale factor if general relativity were
correct.) The right plot shows the corresponding effective
equation-of-state parameter as a function of redshift.}
\label{zplots}
\end{centering}
\end{figure}

An observer, understandably assuming the validity of conventional
general relativity, would interpret measurements of the expansion
history $a(t)$ in MSG in terms of the derived density and dynamics of
an effective dark-energy component $\rho_{\rm DE}^{\rm eff}$.  In terms of
the (in principle) observable quantities $H$ and $\rho$, the
effective dark-energy density is
\be
  \rho_{\rm DE}^{\rm eff} = 3\mpl^2 H^2 - \rho\,.
\ee
where the matter density of course evolves as $\rho\propto a^{-3}$.
The effective equation-of-state parameter is
\be
  w_{\rm eff} = -1 - \frac{1}{3}\dbd{\ln\rho_{\rm DE}^{\rm eff}}{\ln{a}}\,.
\ee
The behavior of these quantities as a function of redshift
in the model defined by (\ref{model2}) is
shown in Figure~\ref{zplots}.

We fix the parameters of our model by fitting the luminosity
distance-redshift relation to the Supernovae Legacy Survey (SNLS) 
data set of 115 type-Ia
supernovae \cite{Astier:2005qq}; in addition we fit for the distance
to the surface of last scattering: we use the third-year WMAP results
\cite{Spergel:2006hy} to fix its redshift, $z_\mathrm{lss} =
1088^{+1}_{-2}$, the acoustic peak scale, $\ell_\mathrm{A} =
302^{+0.9}_{-1.4}$, and its calibration through the matter density,
$\Omega_\mathrm{m} h^2 = 0.1265\pm 0.008$. In doing the fit, we
include both matter and radiation, with a total energy density
$\rho=\rho_m\left(1+\frac{R}{a}\right)$, where $R$ is the radiation
to matter energy density ratio today. In fitting to the data, we
allowed for the existence of spatial curvature, characterized by the
parameter \be
  C \equiv -\frac{3M_{\rm Pl}^2\kappa}{\rho^0_{\rm m}} \, ,
  \label{e:c}
\ee 
where $\rho^0_{\rm m}$ is the matter energy density today. 
The fit prefers a slightly open universe, but achieves
approximately the same $\chi^2$ as $\Lambda$CDM. The set of best fit
parameters is
\begin{equation}
\label{best-fit} \alpha=0.98\nn \qquad C = 0.12\nn \qquad h = 0.72\
.
\end{equation}

Since we are fitting cosmological parameters in our model to the data,
we need a consistent definition of the fractional densities
$\Omega_{\rm m}$ and $\Omega_\kappa$.
To do this, we begin by
defining the critical density today in the usual way, via
$\rho_{\rm cr}^0 = 3M_{\rm Pl}^2 H_0^2$, enabling us
to write the Friedmann equation (\ref{e:fried2}) as
\begin{equation}
  H^2=H_0^2 \frac{16(\alpha a^3 +1)^2}{(\alpha a^3+4)^2}
  \left[\frac{4\left(1+\frac{R}{a}\right)(\alpha a^3 +1) -\alpha a^3 \ln(\alpha a^3)}
  {4a^3\sqrt{\alpha a^3 +1}} +\frac{C}{a^2}\right]
  \frac{\rho^0_{\rm m}}{\rho^0_{\rm cr}} \ .
\end{equation}
Now consider the asymptotic form of this equation as $a\rightarrow 0$.
Ignoring the slowly-varying logarithm, we obtain
\begin{equation}
  H^2=H_0^2 \left[\left(1+\frac{R}{a}\right)\frac{1}{a^3}
  +\frac{C}{a^2}\right]
  \frac{\rho^0_{\rm m}}{\rho^0_{\rm cr}} \ ,
\end{equation}
which then allows us to consistently define the fractional densities
today in precisely the same way as in $\Lambda$CDM:
\begin{equation}
\Omega_{\rm m}=\frac{\rho^0_{\rm m}}{\rho^0_{\rm cr}}\nn =
0.26\qquad \Omega_\kappa=\frac{C\rho^0_{\rm m}}{\rho^0_{\rm cr}} =
+0.02 \,,
\end{equation}
where the numbers are the best-fit values to the supernova and CMB
data.


\section{Linear Perturbations}

To study perturbations, we introduce small time- and space-dependent
deviations from the background cosmological solutions for the metric
and matter sources. Applying the equations of motion and
linearizing, we then obtain a set of coupled first-order
differential equations describing the coupled evolution of matter,
radiation and metric perturbations as the universe expands.

The evolution equations for matter and radiation in the presence of
a perturbed metric are given by the Boltzmann equations (for a
particularly clear description of this see~\cite{Dodelson}). This
formalism is independent of the equations of motion for the metric
itself; moreover, as we work in the matter frame, the Boltzmann
equations are formally the same as those derived in standard general
relativity. However, these equations do
contain new dynamics even if their structure is unchanged,
since the background quantities on which they depend are solutions of
modified background equations.

From (\ref{best-fit}), the best-fit background cosmology requires a
small negative curvature. However, this will negligibly affect
perturbations. Therefore, although we may include curvature in the
background evolution, we neglect it in the treatment of
perturbations. We consider scalar perturbations to a flat FRW metric
in conformal Newtonian gauge; the perturbed line element can be
written as
\begin{equation}
  \ds s^2 = -(1+2\Psi(\vec{x},t))\ds t^2
  + a^2(1+2\Phi(\vec{x},t))\ds{\bf x}^2 \,,
\end{equation}
where $\Phi$ and $\Psi$ are spacetime-dependent gravitational potentials.
We assume that the universe contains both radiation and dark
matter, which allows us to parameterize the perturbations to the
energy momentum tensor by
\begin{eqnarray}
{T^0}_0 &= -\rhobar\delta\\
{T^0}_i &= (1+w)\rhobar\partial_i q \\
{T^i}_j &= w\rhobar({\delta^i}_j+{\pi^i}_j)\,. \label{e:tmunuaniso}
\end{eqnarray}
Here, $\rhobar$ is the background energy density,
$\delta\equiv \delta\rho/\rhobar$, $w$ is the
equation-of-state parameter for the background fluid,
$q$ is the momentum density, and
${\pi^i}_j$ is the anisotropic stress.

We can then linearize the modified Einstein
equations~(\ref{einsteineq0}) and obtain the Poisson and anisotropy
equations for the perturbed cosmology in Fourier space,
\begin{eqnarray}
  &\frac{k^2}{a^2}(\Phi + \delta\psi)
  + 3\left(H+\dot{\psibar}\right)\left(\dot\Phi
    + \dot{\delta\psi}\right)
  - 3\left(H+\dot{\psibar}\right)^2\Psi \nonumber \\
    &\qquad= \frac{\e^{-2\psibar}}{2M_{\rm Pl}} \left( \rhobar\delta +
    (\Ubar'-2\Ubar-2\rhobar)\delta\psi\right) \label{e:poi1}\\
    &\Phi+\Psi =
    -\frac{a^2}{k^2}\frac{\e^{-2\psibar}}{2M_{\rm Pl}^2} w\rhobar\pi
    -2\delta\psi \ ,
    \label{e:aniso}
\end{eqnarray}
where $\psibar$ and $\Ubar$ are the background quantities and
$\delta\psi$ is the perturbation to $\psi$. The scalar anisotropy $\pi$ is
related to the anisotropic stress by ${\pi^i}_j =
\left(\partial^i\partial_j
-\frac{1}{3}\delta^i_j\partial^2\right)\pi$, and is non-zero only for
radiation. We therefore neglect $\pi$ and ${\pi^i}_j$ from now on.

Note that, in comparison with the unmodified ($\Lambda$CDM) version,
the anisotropy equation~(\ref{e:aniso}) contains just one extra
term, namely that involving $\delta\psi$.
However, this term plays a crucial role in defining the
difference between
dark energy and modified gravity. Bertschinger
\cite{Bertschinger:2006aw} has recently emphasized that knowledge of both the
modification to the Friedmann equation and the modification to the
evolution of the anisotropy, $\Phi + \Psi$, is key for the understanding
of the effects of the new gravity theory on cosmological structure
formation. It is the latter which differentiates modified-gravity theories
from models with dark energy, and therefore it is the existence of this
new term which ultimately allows us to test the modified-gravity idea.

In contrast with the anisotropy equation,
the evolution equation (\ref{e:poi1}) has several new
entries arising from the modification to the Einstein equation
(\ref{einsteineq0}). Besides a rescaling of Newton's constant, there
are terms that are functions solely of the scalar field, its
perturbation and the potential. However, by using
equation~(\ref{constraint0}), we can express $\psibar$ and
$\delta\psi$ in terms of the matter density and its perturbation. In
particular, we have
\begin{equation}\label{delta}
    \delta\psi = -\frac{1}{3}\dbd{\psibar}{\ln a} \delta_{\rm m}\ ,
\end{equation}
where $\delta_m$ is just the matter density contrast, since radiation makes
no contribution to $T = -\rho + 3p$.

The time-space $(0,j)$
component of the linearized Einstein equation is
\begin{equation}
\dot{\Phi}-H\Psi=\frac{\e^{-2\psibar}}{2M_P^2}\rhobar (1+w)q
+\dot{\psibar}\Psi-\dot{\delta\psi}+H\delta\psi+\dot{\psibar}\delta\psi
\,.
\end{equation}
Combining this with (\ref{e:poi1}) and making use of~(\ref{delta}),
we obtain an algebraic equation for the potential $\Psi$.
At late times, we are interested in
modes well within the Hubble radius, and we can also neglect contributions
from radiation.  We then have
\begin{equation}
    \frac{k^2}{a^2} \Psi =
    -\left[\frac{\e^{-2\psibar}}{2M_{\rm Pl}^2}\left(1+\dbd{\psibar}{\ln a}
    \right) \rhobar_{\rm m} - \frac{k^2}{3a^2}\dbd{\psibar}{\ln
    a}\right]\delta_{\rm m} \label{e:poi2} \,.
\end{equation}
This constraint relates the potential $\Psi$ directly to the matter
variables.


\section{The Growth of Linear Structure}

In the previous section we have derived the basic linearized
equations for the evolution of perturbations in the context of
modified-source gravity. In this section we study the growth of
linear structure.

At late times, when all the modes of interest have entered the
horizon, and radiation and momentum flow are negligible, we can
combine (\ref{e:poi2}) with the Boltzmann equations for dark matter
to obtain a second-order differential equation governing dark matter
perturbations
\begin{equation}
    \ddot{\delta}_{\rm m} + 2H\dot{\delta}_{\rm m}
  - \left[\frac{\e^{-2\psibar}}{2M_{\rm Pl}^2}\left(1+\dbd{\psibar}{\ln a}
    \right) \rhobar_{\rm m} - \frac{k^2}{3a^2}\dbd{\psibar}{\ln
    a}\right]\delta_{\rm m}= 0 \label{e:gro} \,.
\end{equation}
The first term in the bracket multiplying $\delta_{\rm m}$ is the
same as the term in the Poisson equation in GR, modified by the fact
that the value of Newton's constant is evolving as the average
density in the universe decreases beneath the critical value at
which the modifications to GR become important.

\begin{figure}[t]\begin{centering}
\includegraphics[width=.7\columnwidth]{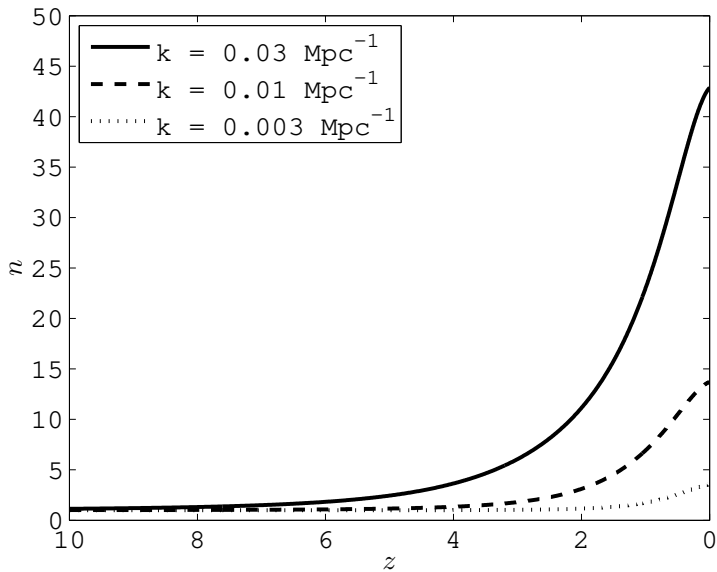}
\caption{The exponent of the growth of matter density perturbations,
as defined in~(\ref{e:ge}) for a selection of modes, solved for the
model defined by (\ref{model2}) with the best fit parameters
(\ref{best-fit})(near-horizon effects are ignored). The rate of
growth is very strongly scale-dependent once the cosmology enters
the epoch of acceleration, with the exponent departing sharply from
$n=1$ and tending to $n \propto k$ at late times. \label{f:groexp}}
\end{centering}\end{figure}

The second term in the coefficient of $\delta_{\rm m}$ in
(\ref{e:gro}) is very different in nature: it introduces a scale
dependence in the growth of structure.  For negative
$\ds\psibar/\ds\ln a$, small scales will begin to grow more quickly
than large scales once the universe approaches the accelerating
phase. In figure~\ref{f:groexp}, we give an example of this enhanced
growth for the choice of potential corresponding to the model
defined by (\ref{model2}) with the best fit parameters
(\ref{best-fit}). We have plotted the exponent $n$ of
\begin{equation}
    \delta_{\rm m} \equiv \delta_0
  \exp\left[\int \ds\!\ln a \,n(a)\right] \label{e:ge}\ ,
\end{equation}
with $\delta_0$ a constant. The rate of growth is
strongly scale-dependent once the universe enters the epoch of
acceleration, with the exponent
departing from $n=1$ (holding for all modes during matter
domination) and tending to $n\propto k$ at late times.
The $k$-dependence of the growth rate will result in smaller scales
reaching non-linearity extremely quickly after the onset of
acceleration. Thus, we generically expect that the matter power
spectrum will be non-linear at larger scales than in $\Lambda$CDM.

In fact, this enhanced growth of scales which are small but still in
the linear regime seems to be a generic feature of modified-source
gravity.  The effect can be traced to the existence of the new
$k$-dependent term in equation (\ref{e:ge}) for $\delta_{\rm m}$:
small scales will witness enhanced growth so long as the potential
$U(\psi)$ is chosen so that $\ds\psibar/\ds\ln a < 0$. But, as
inspection of (\ref{e:fried}) shows, it is precisely this behavior
that makes the universe accelerate at late times, by increasing the
effective value of Newton's constant in the modified Friedmann
equation.  It therefore seems difficult to avoid this phenomenon
simply by a clever choice of the potential $U(\psi)$.

A possible loophole in this argument would be to consider models in
which $\psi$ were nearly constant in the present era, but the
contribution of the potential $U(\psi)$ itself to the right-hand side
of (\ref{e:fried}) were to induce cosmological acceleration.
One might reasonably complain that this case is simply a dark-energy
model in disguise, rather than a modification of gravity; however, we
should keep in mind that $U(\psi)$ is not really the potential for a
dynamical scalar field, but rather a nonlinear function of the matter
variables.  This is something of a matter of taste, and we will not
pursue this possibility in the remainder of the paper.

\begin{figure}[t]\begin{centering}
\includegraphics[width=.7\columnwidth]{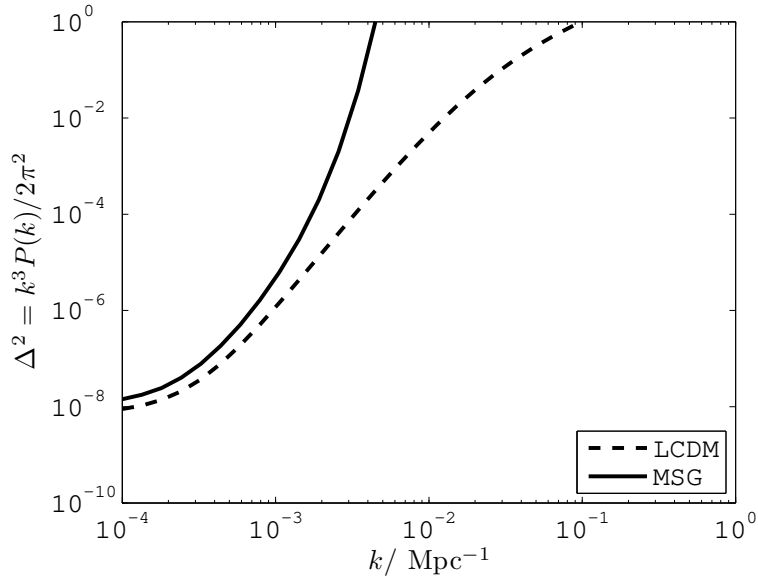}
\caption{The dimensionless matter power spectrum for MSG and
$\Lambda$CDM. As expected, in MSG the growth-rate is proportional to
$k$ and the power-spectrum increases exponentially with $k$ for
modes within the horizon. Non-linear scales are reached at scales of
$240$ Mpc ($k=0.004$~Mpc$^{-1}$), compared to
$10$~Mpc$\equiv8h^{-1}$ Mpc for $\Lambda$CDM. The linear growth can
only be trusted for $k<0.004$~Mpc$^{-1}$. For smaller scales, the
theory should return to a GR-like behavior.} \label{f:matpow}
\end{centering}\end{figure}

We proceed to study structure formation in this model. We start with
a Harrison-Zel'dovich scale-invariant spectrum of perturbations
normalized to COBE by $\Delta^2_\zeta = 5.07 \times 10^{-5}$, (where
$\Delta$ is the dimensionless square root of the matter power
spectrum defined as $\Delta^2\equiv k^3 P(k)/2\pi^2$), and evolve through to today using a
numerical code that includes both radiation and dark matter. Since
we expect small scales to be non-linear, it is necessary to find the
scale at which linear perturbation theory ceases to be valid. These
results are shown in figure~\ref{f:matpow}.

As expected, the non-linear terms are important at much larger
scales than in $\Lambda$CDM: approximately $240$~Mpc compared to
$10$~Mpc for $\Lambda$CDM. Once the perturbations are non-linear,
perturbation theory breaks down and an N-body simulation with
potentials depending on local densities must be used to obtain
results.  (Nonlinearities become important when $\Delta\sim 1$; the
potentials $\Phi$ and $\Psi$ are safely smaller at that time.)
However, since for high-enough densities the equations of motion
return to GR, we expect that the cosmology should behave in a
GR-like manner for sufficiently high values of $\delta$.

\begin{figure}[t]
\begin{centering}
\includegraphics[width=.7\columnwidth]{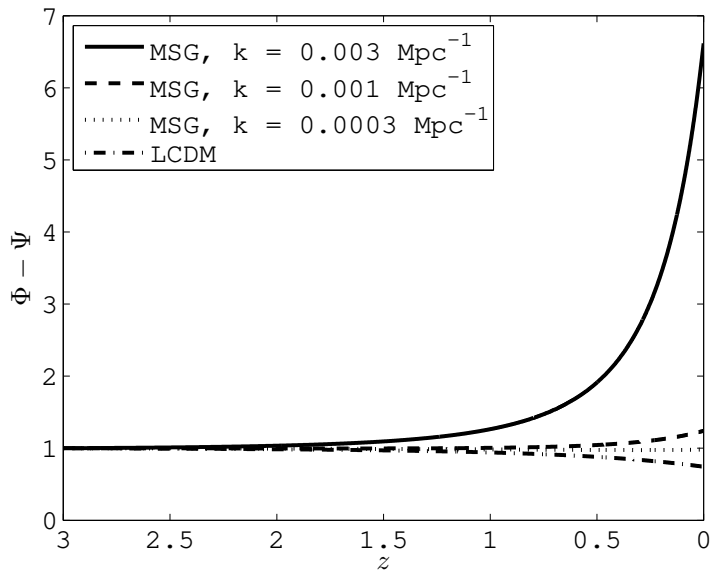}
\caption{The evolution of the quantity $\Phi-\Psi$---observable
through the ISW effect---for modes close enough to the horizon to
remain linear until today. In MSG, this combination of potentials
grows during dark-energy domination, increasingly rapidly for higher
$k$. This would lead to a significant increase in the ISW signal, at
least for the lowest multipoles, which are sensitive only to the
linear modes at largest scales. In LCDM this evolution is scale
free.\label{f:phi}}
\end{centering}
\end{figure}

\begin{figure}[t]
\begin{centering}
\includegraphics[width=.7\columnwidth]{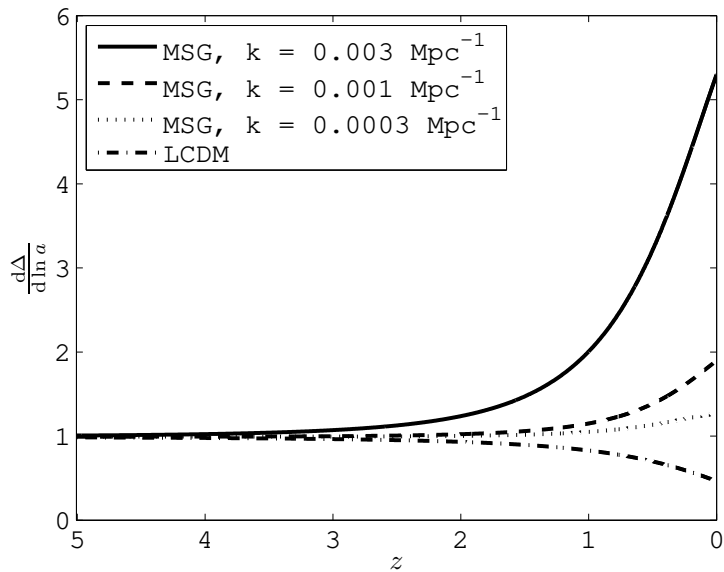}
\caption{Exponent of growth rate for comoving density contrast,
\dbd{\Delta}{\ln a}. In $\Lambda$CDM, it is scale-independent once
radiation no longer dominates; structure growth slows down during
dark-energy domination. For MSG, structure growth accelerates.}
\label{f:delta}
\end{centering}
\end{figure}

The evolution of modes at very large scales, for which linear
perturbation theory is valid throughout the history of the universe,
is presented in figures~\ref{f:phi} and~\ref{f:delta}. Structure
growth is scale dependent and runaway, as predicted by the
approximation in~(\ref{e:gro}), with the rate of growth increasing
as the universe departs from conventional matter-dominated behavior.
Such rapid structure formation drives the growth of gravitational
potentials, which also increase rapidly during the acceleration era
in a scale-dependent manner. This behavior would significantly
enhance the Integrated Sachs-Wolfe (ISW) effect, at least at the
lowest multipoles which are sensitive to only the largest scales,
which remain linear. In addition, the fact that the gravitational
potentials are growing at the same time as the density contrasts are
increasing would lead to a galaxy-ISW correlation of opposite sign
to that expected in $\Lambda$CDM; whereas these are correlated for
$\Lambda$CDM, here they would be anti-correlated.

\section{Conclusions}

We have considered a class of modified-gravity models in which only 
the constraint equation of GR is modified, thereby introducing no new
propagating degrees of freedom. These ``Modified-Source Gravity" models 
allow for cosmological dynamics in which the universe self-accelerates, 
without the need for dark energy. We demonstrate that there exists a 
class of such theories with a consistent cosmic expansion history and 
which naturally satisfies all solar system tests of gravity.

As with any explanation for cosmic acceleration, it is important to 
understand how it might be tested and distinguished from other models. 
To this end, we have studied the
onset of structure formation in modified-source
gravity using linear perturbation
theory. Fixing parameters so that the background cosmology is well
described by MSG, we have compared
the rate of growth of different Fourier modes of the density
perturbations with those predicted in the
$\Lambda$CDM model. We find that, for a given $k$-mode,
growth is more rapid in MSG than in
$\Lambda$CDM, and therefore that perturbation theory breaks down at
a correspondingly higher redshift.

To make detailed progress beyond this point would require an N-body
simulation with potentials depending on local densities. However,
since MSG is constructed so as to yield dynamics indistinguishable
from GR at high enough densities, we expect this rapid growth to
cease and to once again resemble that found in $\Lambda$CDM for
sufficiently high values of $\delta_{\rm m}$.

Particularly interesting is the evolution of modes on scales large
enough that linear perturbation theory is valid throughout the history
of the universe. The rapid structure formation on these scales drives
the growth of gravitational potentials, which increase rapidly during
the acceleration era in a scale-dependent manner. This behavior
enhances the Integrated Sachs-Wolfe effect at the lowest
multipoles. Since the gravitational potentials are growing at the same
time as the density contrasts are increasing this should lead to a
galaxy-ISW anti-correlation, in contrast with that expected in
$\Lambda$CDM.  A natural next step is to attempt to understand the
growth of structure in the nonlinear regime.

A primary motivation for this work has been to understand the way in
which modified gravity can be distinguished from dark energy.
Currently, the leading candidate for a modified theory of
cosmological gravity is the DGP model
\cite{Dvali:2000hr,Deffayet:2000uy,Deffayet:2001pu}, despite
lingering fundamental issues with the theory
\cite{Luty:2003vm,Adams:2006sv}. Much effort has gone into
understanding the evolution of cosmological perturbations in DGP
gravity, with an emerging consensus that the gravitational
potentials decay more rapidly at late times in DGP than they do in
$\Lambda$CDM \cite{Lue:2004rj,Knox:2005rg,Lue:2005ya,
Sawicki:2005cc,Koyama:2005kd,Koyama:2006ef,Song:2006sa,Sawicki:2006jj,
Song:2006jk}.  In MSG, in contrast, it appears as if the potentials 
generically \emph{grow} at late times with respect to their conventional
behavior.  It therefore seems to be difficult to imagine a
model-independent prediction for the way in which modified gravity
can be distinguished from theories of dynamical dark energy,
although simultaneous measurements on the expansion history and the
evolution of structure do of course provide stringent constraints on
any specific model.  It is clearly important to continue to explore
the theoretical consequences of modifying general relativity on
large scales, to better understand what clues observers should be
looking for in the quest to solve the puzzle of the accelerating
universe.

\section*{Acknowledgments}
The work of SC and IS is supported in part by U.S.~Dept.~of Energy
contract DE-FG02-90ER-40560, NSF grant PHY-0114422 (KICP) and the
David and Lucile Packard Foundation. The work of AS and MT is
supported in part by the NSF under grant PHY-0354990, by Research
Corporation, and by funds provided by Syracuse University. SC thanks
Nima Arkani-Hamed, Lawrence Yaffe, and Matt Strassler for helpful
discussions, and the Green Mill for generous hospitality. MT and AS
thank Levon Pogosian for useful discussions.

\end{document}